\documentstyle[PASJadd]{PASJ95}
\begin{document}
\title{J-Net Galactic Plane Survey of VLBI Radio Sources \\
for VLBI Exploration of Radio Astrometry (VERA)}
\author{Mareki {\sc Honma}$^{1,2}$, Tomoaki {\sc Oyama}$^{1,3}$, Kazuya {\sc Hachisuka}$^{1,5}$, Satoko {\sc Sawada-Satoh}$^{1,2}$, Kouichi {\sc Sebata}$^{1,9}$,\\
Makoto {\sc Miyoshi}$^{1,4}$, Osamu {\sc Kameya}$^{1,2}$, Seiji {\sc Manabe}$^{1,2,4}$, Noriyuki {\sc Kawaguchi}$^{1,4}$, Tetsuo {\sc Sasao}$^{1,4}$,\\
Seiji {\sc Kameno}$^6$, Kenta {\sc Fujisawa}$^6$, Katsunori M. {\sc Shibata}$^{4,6}$, Takeshi {\sc Bushimata}$^6$, Takeshi {\sc Miyaji}$^{1,7}$,\\
Hideyuki {\sc Kobayashi}$^6$, Makoto {\sc Inoue}$^{6,7}$, Hiroshi {\sc Imai}$^{1,2,4}$, Hiroshi {\sc Araki}$^2$, Hideo {\sc Hanada}$^4$,\\
Kenzaburo {\sc Iwadate}$^2$, Yoshihisa {\sc Kaneko}$^4$, Seisuke {\sc Kuji}$^4$, Katsuhisa {\sc Sato}$^2$, Seiitsu {\sc Tsuruta}$^2$,\\
Satoshi {\sc Sakai}$^4$, Yoshiaki {\sc Tamura}$^4$, Koji {\sc Horiai}$^2$, Tadayoshi {\sc Hara}$^4$, Koichi {\sc Yokoyama}$^4$,\\
Junichi {\sc Nakajima}$^8$, Eiji {\sc Kawai}$^8$, Hiroshi {\sc Okubo}$^8$, Hiro {\sc Osaki}$^8$, Yasuhiro {\sc Koyama}$^8$,\\
Mamoru {\sc Sekido}$^8$, Tomonari {\sc Suzuyama}$^8$, Ryuichi {\sc Ichikawa}$^8$, Tetsuro {\sc Kondo}$^8$, Kensei {\sc Sakai}$^{10}$, \\
Katsuhiro {\sc Wada}$^{10}$, Naohiko {\sc Harada}$^{10}$, Norihiro {\sc Tougou}$^{10}$, Mitsumi {\sc Fujishita}$^{10}$,\\
Rie {\sc Shimizu}$^{11}$, Satomi {\sc Kawaguchi}$^{11}$, Akane {\sc Yoshimura}$^{11}$, Masakazu {\sc Nakamura}$^{11}$, Wataru {\sc Hasegawa}$^{11}$,\\
Satoshi {\sc Morisaki}$^{11}$, Ryuichi {\sc Kamohara}$^{11}$, Tomoe {\sc Funaki}$^{11}$, Naoko {\sc Yamashita}$^{11}$, Teruhiko {\sc Watanabe}$^{11}$,\\
Tomomi {\sc Shimoikura}$^{11}$, Masanori {\sc Nishio}$^{11}$, Toshihiro {\sc Omodaka}$^{11}$, Atuya {\sc Okudaira}$^{12}$\\
\\
$^1${\it VERA Project Office, NAOJ, Mitaka, Tokyo, 181-8588}\\
$^2${\it Mizusawa Astrogeodynamics Observatory, NAOJ, Mizusawa, Iwate 023-0861}\\
$^3${\it Department of Astronomy, University of Tokyo, Bunkyou, Tokyo 113-8654}\\
$^4${\it Earth Rotation Division, NAOJ, Hoshigaoka, Mizusawa, Iawate 023-0861}\\
$^5${\it Graduate University for Advanced Studies, Hayama, Kanagawa 240-01}\\
$^6${\it VSOP Project, NAOJ, Mitaka, Tokyo, 181-8588}\\
$^7${\it Nobeyama Radio Observatory, NAOJ, Minamisaku, Nagano 384-1305}\\
$^8${\it Kashima Space Research Center, Communications Research Laboratory, Kashima, Ibaraki 314-0012}\\
$^9${\it Communications Research Laboratory, Koganaei, Tokyo 184-8795}\\
$^{10}${\it School of Engineering, Kyushu-Tokai Univerisity, Kumamoto 862-8652}\\
$^{11}${\it Faculty of Science, Kagoshima University, Kagoshima 890-0065}\\ 
$^{12}${\it International University of Kagoshima, Kagoshima 891-0191}\\
{\it E-mail(MH): honmamr@cc.nao.ac.jp}}

\abst{
In order to search for new VLBI sources in the Galactic plane that can be used as phase reference sources in differential VLBI, we have conducted 22 GHz observations of radio sources in the Galactic plane using the Japanese VLBI Network (J-Net).
We have observed 267 VLBI source candidates selected from existing radio surveys and have detected 93 sources at the signal-to-noise ratio larger than 5.
While 42 of the 93 detected sources had already been detected with VLBI at relatively lower frequency (typically 2 to 8 GHz), the remaining 51 are found to be new VLBI sources detected for the first time.
These VLBI sources are located within $|b|\le 5^\circ$, and have a large number of Galactic maser sources around them.
Thus, they are potential candidates for phase reference sources for VLBI Exploration of Radio Astrometry (VERA), which is the first VLBI array dedicated to the phase referencing VLBI astrometry aiming at measuring the parallax and proper motion of maser sources in the whole Galaxy.
}
\kword{radio sources --- radio galaxies --- QSOs --- VERA}
\maketitle
\thispagestyle{headings}

\section{Introduction}
VLBI Exploration of Radio Astrometry (VERA), approved to start its construction in 2000, is a new Japanese VLBI array of 20m antennas dedicated to astrometry (Honma et al. 2000; Kawaguchi et al. 2000).
VERA will undertake astrometry of Galactic maser sources based on the differential VLBI technique, in which two adjacent sources are observed simultaneously to remove the atmospheric fluctuation.
A position accuracy of $\sim 100$ $\mu$arcsec has been already achieved by switching differential VLBI in which an object and a reference are observed by turns (e.g., Lestrade et al. 1999), but the atmospheric fluctuation is not completely suppressed because of the time lag between observations of an object and a reference.
In order to obtain a higher position accuracy, one needs to observe an object and a reference source at the same time.
VERA, with its dual beam antenna system, will enable us to observe a Galactic maser and an extra-galactic reference source simultaneously.
With such a system, positions of maser sources relative to reference sources can be measured with 10 $\mu$arcsec level accuracy, allowing us to obtain parallaxes of maser sources in the whole Galaxy.

To remove the atmospheric fluctuation effectively, a reference source must be sufficiently close to the maser source for which one intends to measure the parallax and proper motion.
In order for VERA to achieve a 10 $\mu$arcsec level accuracy, it is required that a maser and a reference source must be located within 2$^\circ$ from each other.
This requires a large number of reference sources spreading over the whole sky area.
Currently, there exist several catalogs that contain hundreds of continuum sources detected with VLBI (e.g., Preston et al. 1985; Morabito et al. 1986; Johnston et al. 1995; Ma et al. 1998; Peck \& Beasley 1998).
In total, more than 2000 VLBI sources are known.
However, the number of VLBI sources is considerably smaller in the Galactic plane than that in the off-plane region.
This is because several surveys of VLBI sources were conducted in the off-plane regions:
for example, the VLBI survey by Preston et al.(1985) are for $|b|\ge 10^\circ$, and the VLBA calibrator survey by Peck \& Beasley (1998) are mainly based on the Jodrell Bank-VLA Astrometric Survey (Patnaik et al. 1992; Browne et al. 1998; Wilkinson et al. 1998), which are conducted at $|b|\ge 2.5^\circ$.
On the other hand, H$_2$O and SiO masers, which are the main targets of VERA, are mostly concentrated in the Galactic plane because they are emitted from Galactic star forming regions and Mira-type variable stars: for example, about 40\% of H$_2$O maser sources listed in Palagi et al.(1993) are located within $b=\pm 5^\circ$.
Therefore, a large number of VLBI sources are necessary as phase and position reference for VERA, and hence the search for VLBI sources in the Galactic plane is indispensable to the success of VERA.

A search for VLBI sources in the Galactic plane has its own scientific merits even apart from the necessity for VERA.
First, many of the continuum radio sources are likely to be radio galaxies and QSOs, and so a search for radio sources in the Galactic plane will provide a basis for studying the distribution of galaxies and large scale structures behind the Milky Way.
Second, a search for VLBI sources in the Galactic plane may lead us to discoveries of new active Galactic sources like X-ray binaries or flare stars which are of significant interest in stellar physics and high energy astrophysics.
Third, the radio sources can be used to study the cold interstellar matter in the Galaxy by observing the absorption line systems toward them, and also can be used to investigate the ionized interstellar matter by evaluating the interstellar scintillation effect on these sources.
Hence, a search for radio sources in the Galactic plane is useful not only for VERA, but also for a number of interesting studies in astronomy and astrophysics.
For these reasons, we have performed a survey of continuum radio sources in the Galactic plane using the Japanese VLBI Network (hereafter J-Net; Kameya et al. 1993; Shibata et al. 1994; Omodaka et al. 1994) at 22GHz, and here we report on the results.

\section{Sample Selection}

In order to find VLBI sources effectively, we selected VLBI source candidates from existing radio surveys.
The source candidates in the northern sky ($\delta\ge 0^\circ$) were selected from the single-dish sky survey made at Green Bank (Becker et al. 1991, hereafter Green Bank survey) and the low resolution interferometer survey made with the Texas interferometer (Douglas et al. 1996, hereafter Texas survey).
The northern VLBI source candidates were selected under the following selection criteria:
1) sources that are listed both in Green Bank survey and Texas survey,
2) sources that are located at $|b|\le 5^\circ$,
3) sources that have an expected flux density at 22 GHz larger than 100 mJy (calculated from flux density at 5 GHz in Green Bank survey and at 0.3GHz in Texas survey), and 
4) sources that are not classified as an extended source in the Green Bank survey.
The last condition was introduced because highly extended sources are likely to be resolved out when observed with VLBI.
The number of VLBI source candidates selected under these criteria is 198.

To examine whether the sources selected under such criteria are likely to be VLBI sources, we applied the criteria 1), 3) and 4) to the radio sources located at $b\ge 30^\circ$, where a large number of VLBI sources are listed in the International Celestial Reference Frame catalog (Ma et al. 1998, hereafter ICRF catalog) and VLBA calibrator survey (Peck \& Beasley 1998).
We confirmed that more than 40\% of sources which are at $b\ge 30^\circ$ and satisfy the criteria 1), 3) and 4) are indeed known VLBI sources listed in ICRF catalog or VLBA calibrator survey.

Since part of the southern sky is also observable with J-Net and VERA, VLBI source candidates were also selected in the southern Galactic plane.
Similar to the northern sample, the southern sample sources were selected based on the following criteria:
1) sources that are listed both in Parkes-MIT-NRAO survey (Griffith et al. 1994; 1995, hereafter PMN survey) and Texas survey,
2) sources that are located in the southern Galactic plane observable with J-Net ($|b|\le 5^\circ$, $0^\circ \le l\le 33^\circ$ or $213^\circ\le l \le 243^\circ$),
3) sources that have a flux larger than 200 mJy at 5 GHz, and 
4) sources that are not classified as an extended source in the PMN survey.
The total number of the sources which satisfy the criteria above are 86.
Including 198 northern VLBI source candidates, our sample consists of 284 continuum sources located at $0^\circ\le l\le 243^\circ$ and $-5^\circ\le b\le 5^\circ$.
Table 1 lists the 284 VLBI source candidates with positions taken from Texas survey.
As discussed later, some of the sources in table 1 are already known as VLBI sources.
Nevertheless, they were included in the source list and observed with J-Net because measurements of their correlated flux density at 22 GHz are necessary for VERA.
For known VLBI sources, we present in table 1 the most accurate positions taken from ICRF catalog and VLBA calibrator survey.

\section{Observation and Data Reduction}

The observation was carried out at 22 GHz during Oct 7 - 11, 1999 (100 hours in total) using J-Net, which consists of four stations; Nobeyama 45m, Kashima 34m, Mizusawa 10m and Kagoshima 6m.
Since the main purpose of this observation is to detect as many sources as possible, integration time was set to be quite short, typically 5 minutes.
This integration time is longer than a typical VLBI coherence time at 22 GHz ($\sim 100$ sec), and thus sufficient for fringe search.
In order not to miss some detections due to bad weather or system trouble, most of sources were observed twice or more.
Strong H$_2$O maser sources, which were used to check the system performance and to calibrate the flux of continuum sources, were also observed approximately every two hours with an integration time of 5 minutes.
Known VLBI sources were also observed frequently to calibrate clock offsets in the hydrogen-maser frequency standard at each station.
Signals were sampled at the rate of 32$\times 10^6$ sample per second with 2-bit quantization, and recorded at the rate of 128 Mbps using the VSOP terminal with two baseband channels of 16 MHz bandwidth.
In order to avoid detecting H$_2$O masers at the band edge, the frequency coverage of the two channels were set to overlap slightly, covering from 22.220 GHz to 22.250 GHz in total.
The correlation processing was carried out with the Mitaka FX correlator (Chikada et al. 1991) located at NAOJ Mitaka campus, and fringe search was made using Global Fringe Search (GFS) procedure.
The size of the fringe search window was 1 $\mu$sec by 20 psec/sec in residual delay and delay rate, respectively.
Fringe search was mainly done for the Nobeyama-Kashima baseline, which has the shortest baseline and highest sensitivity among six J-Net baselines, and other baselines were used supplementary, in particular for position measurements (see section 4.3).

\section{Results}

\subsection{Fringe Detection}

Out of 284 sources listed in table 1, we observed 267 sources, including 51 previously-known VLBI sources that are listed in any of ICRF catalog (Ma et al. 1998), VLBA calibrator survey (Peck \& Beasley 1998) or 2 GHz VLBI survey (Preston et al. 1985; Morabito et al. 1986).
Out of 267 observed sources, 93 sources were detected with Nobeyama-Kashima baseline at the signal-to-noise ratio (S/N) larger than 5.
This detection criterion corresponds to a confidence level of 99.5\%.
The S/N for Nobeyama-Kashima baseline is listed in table 1.
The total detection rate of the sources listed in table 1 is about 35 \% (=93/267).
This relatively high detection rate suggests that the simple selection criteria described in section 2 work quite effectively as selection criteria for VLBI sources.
Hence, such selection criteria may be useful for further searches for VERA's reference sources at any Galactic latitude.
We note that the detection limit of VERA array is not so different from that of J-Net Nobeyama-Kashima baseline.
While the product of aperture diameter for Nobeyama-Kashima baseline is about four times larger than that of a VERA baseline, the recording rate of VERA array will be 1G bit per second, which is eight times higher than that of J-Net.
Thus, the detection limit of Nobeyama-Kashima baseline is lower than that of a VERA baseline by only a factor of 1.35.

Among 93 sources detected in the present study, 42 were known VLBI sources.
Since the number of observed known VLBI sources was 51, the 22 GHz detection rate of known VLBI sources is 82\%.
Note that these known VLBI sources were mostly discovered at relatively low frequency (typically from 2 to 8 GHz) and selected based on the flux estimates described in section 2.
The high detection rate in the present study indicates that our flux estimates based on the low resolution observations worked relatively well.

As for the remaining 51 detected sources, detections were not reported in any literature.
Thus, as far as we know, these 51 sources were detected with VLBI for the first time.
Figure 1 shows the distribution of the new VLBI sources in Galactic coordinates as well as other sources listed in table 1.
From figure 1 one can see that the newly detected VLBI sources are distributed almost randomly in the Galactic plane within $|b|\le 5^\circ$ and $0^\circ\le l \le 243^\circ$.
However, the detection rate at the Galactic center region is smaller than that in the other regions.
Possible reasons for this are: 1) maximum elevations of the sources at the Galactic center region are quite low, and thus relatively high system noise temperature makes it difficult to detect them;
2) the high column density of interstellar plasma toward the Galactic center region causes strong interstellar scintillation and thus the radio sources are resolved out when observed with VLBI;
and/or 3) since star formation activity is significantly higher near the Galactic center, many of the VLBI source candidates at the Galactic center region may be compact HII regions that are resolved out with VLBI.

\subsection{Flux Measurement}

For 93 detected sources, we obtained the correlated flux density for the Nobeyama-Kashima baseline.
The correlated flux density was calculated using the following relation,
\begin{equation}
S=\frac{2k\rho}{\eta_c}\frac{\sqrt{T_{\rm sys1}T_{\rm sys2}}}{\sqrt{{\eta_{e1}\eta_{e2}A_1 A_2}}}.
\end{equation}
Here $\rho$ is the correlation coefficient, $T_{\rm sys}$ is the system noise temperature, $A$ is the antenna aperture, $\eta_e$ is the antenna efficiency, $k$ is the Boltzmann constant, and $\eta_c$ is the sampling efficiency which equals to 0.88 for the 2-bit sampling mode.
The antenna aperture of Nobeyama and Kashima stations are 1590 m$^2$ (45m$\phi$) and 908 m$^2$ (34m$\phi$), and the aperture efficiencies are $0.60$ and $0.57$, respectively.
The correlation coefficient $\rho$ was calculated with GFS procedure by integrating the observational data within the coherence loss limit.
Typical timescale of the coherence loss was found to be $\sim$120 seconds.
The system noise temperature $T_{\rm sys}$ was monitored roughly every two hours at both Kashima and Nobeyama stations based on the R-Sky method, in which a blank sky and a reference black body were observed.
To correct for the elevation differences for each source, we converted $T_{\rm sys}$ into the optical depth at the zenith $\tau_0$, which relates to $T_{\rm sys}$ through the following equation,
\begin{equation}
T_{\rm sys} \exp(-\tau_0 \sec z) = T_{\rm RX}+T_{\rm atm}\left(1-\exp(-\tau_0 \sec z)\right).
\end{equation}
Here $T_{\rm RX}$ is the receiver noise temperature, $T_{\rm atm}$ is the atmospheric temperature, and $z$ is the zenith distance.
Note that a factor of $\exp(-\tau_0 \sec z)$ in the left-hand side corrects for the atmospheric attenuation.
The atmospheric temperature $T_{\rm atm}$ at both stations is assumed to be 290K, which is accurate within 3\% during the observation.
The receiver temperature at the Kashima station was 240K with its variation of $\sim$ 3\% during the observation period.
The receiver temperature at Nobeyama station was estimated to be 100K from the $T_{\rm sys}$ dependence on the $\sec z$, and assumed to be constant during the observation.

The correlated flux densities for 93 sources were calculated from equation (1) and are listed in table 1.
Since all detected sources have S/N ratio larger than 5, the error in flux density caused by the correlation coefficient measurement is less than 20\%.
However, the variation of atmospheric condition with timescale shorter than two hours (the timescale of $T_{\rm sys}$ measurement) is another cause for errors in flux density measurement, and thus the typical error of flux density in table 1 is not much less than $\sim$ 20\%.
Figure 2 shows the number distribution of correlated flux densities.
As seen in figure 2, the number of sources increases in a power law form with decreasing the flux density, as usually expected.
Figure 2 also demonstrates that while bright sources were mostly detected with previous observations, many of faint sources are detected for the first time by our observation.

\subsection{Position measurement}

Some of the sources we observed were detected with multiple baselines.
For the sources which are detected with 3 or more independent baselines, we estimated the source position using the delay rate.
The delay rate and the source position are related to each other through following equation:
\begin{equation}
c \Delta\dot{\tau}_g = \dot{U} \Delta x+ \dot{V} \Delta y,
\end{equation}
where $c$ is the speed of light, $U$ and $V$ are physical baseline length projected onto the u-v plane, and the dot indicates the time derivative.
The observable $\Delta\dot{\tau}_g$ is defined as the difference between the observed geometric delay rate and the predicted geometric delay rate calculated from the given source positions and baseline parameters.
This difference in the delay rate is caused by the position offset $\Delta x$ and $\Delta y$, which is defined as the difference between the position of phase tracking center (position given in table 1) and the real source.
Note that $\Delta x$ is measured in the direction of right ascension and $\Delta y$ is measured in the direction of declination, as commonly done.
The geometric delay rate difference $\Delta\dot{\tau}_g$ as well as the time derivative of baseline components $U$ and $V$ were calculated with GFS procedure.

In table 2, we list the positions for 21 sources that are detected with 3 or more baselines.
Note that all of 21 sources are new VLBI sources except TXS0601+245, which was detected by Morabito et al.(1986), but accurate position was not reported.
Because of short observation (typically 5 minutes for each source) and relatively short baselines (for example, Nobeyama-Kashima baseline is 197 km, and Nobeyama-Mizusawa baseline is 425 km), the positional accuracy is of $\sim$ 1 arcsec.
As seen in table 2, the positions obtained with Texas survey (listed in table 1) are usually accurate within a few arc-seconds, but there are some sources showing position errors larger than 10 arc-seconds.
Thus, one has to be careful in search for an optical counterpart of Texas sources.

\section{Discussion}

As seen in previous sections, we have detected 93 of 267 Texas survey sources that are selected under simple criteria, and found 51 new VLBI sources in the Galactic plane.
The total detection rate of 35\%(=93/267) is quite high, and thus one can use the same criteria to search for more VLBI sources in the other sky area.
Also, the 22 GHz detection rate of known VLBI sources that were previously detected at relatively low frequency (typically 2 to 8 GHz) are quite close to unity.

Since the detection limit of J-Net is quite close to that of VERA, most of the sources detected in the present study are also observable with VERA.
Thus, one can expect that the discovery of 51 new VLBI sources increases significantly the number of maser-QSO pairs that can be observed with VERA.
In fact, there are 129 H$_2$O maser sources within 2$^\circ$ of 51 new VLBI sources, according to the H$_2$O maser catalog by Palagi et al.(1993).
This number roughly corresponds to 1/6 of independent H$_2$O maser sources in Palagi et al.(1993).
Thus, the sources detected in this study will be of great use as reference sources for VERA.

In order to use the radio sources discovered in the present study (and also other sources) as reference for differential VLBI, one has to know the nature of sources, particularly whether the source is a Galactic or extra-Galactic object.
If the source is a Galactic object, then its parallax and proper motion are not negligible, being unsuitable for position reference.
The best way to know the source nature is optical-identification and spectroscopic measurement of source redshifts.
Unfortunately, optical counterparts and their redshifts are known for only few sources in table 1.
This is partly because positions obtained with single dish and low resolution interferometer observations are not accurate enough for optical identification, and partly because the sources in the Galactic plane suffer significantly from Galactic extinction at the optical band.

The problem of position accuracy can be resolved by measuring its position with VLBI at an accuracy of 1 arcsec or less.
As described in section 4, accurate portions of 21 sources that were detected with 3 or more baselines are obtained in the present study.
In the Digital Sky Survey images, we have found 7 possible optical counterparts, but redshifts are given for only 1 of them in the NASA Extra-galactic Database.
Thus, spectroscopic observation of these optical counterparts are required to understand the source nature.

The fact that possible optical counterparts are found for about 1/3 of the sources with accurate positions is encouraging for further study of the nature of these radio sources in the Galactic plane.
If deep observations are carried out at optical or infrared wavelengths, one may be able to find optical counterparts for a large number of VLBI sources in the Galactic plane, provided that accurate positions are available.
Therefore, the next step to understand their nature is to measure source positions at 1 arcsec level, and this can be easily achieved using current VLBI technique.

\section{Conclusion}

We have searched for new VLBI sources in the Galactic plane using J-Net.
We have observed 267 sources that are selected from single dish and low resolution interferometer observations, and detected 93 sources including 51 sources that are detected with VLBI for the first time.
We have presented the correlated flux for Nobeyama-Kashima baseline for all sources we detected, and also positions for 21 sources that are detected with 3 or more baselines.
These VLBI sources are located within $|b|\le 5^\circ$ of the Galactic plane, and there are a large number of maser sources within 2$^\circ$ of them.
Thus, the sources detected in the present study are potential candidates for the reference sources of phase-referencing VLBI astrometry with VERA.

\section*{acknowledgment}

We are grateful to the NRO staffs, especially Drs. N.Ukita, K.Sunada and N.Nakai, for kindly providing us the observation time of the 45m telescope as a part of their supports to the VERA project.
We also thank the referee for careful reading the manuscript and several constructive suggestions.
Parts of data reduction in the present study were carried out on the workstations at Astronomical Data Analysis Center of the National Astronomical Observatory, Japan.

\section*{figure caption}
\def\re{\hangindent=1pc \noindent}

\re Figure 1 : Distribution of observed radio sources in Galactic coordinates. Filled circles are 51 new VLBI sources, and open circles are 42 previously known VLBI sources that are also detected at 22 GHz in the present study.
Crosses are the radio sources that were observed but not detected.

\re Figure 2 : Distribution of correlated flux density at 22 GHz for Nobeyama-Kashima baseline. Filled area shows distribution of correlated flux density for previously known VLBI sources.

\clearpage
\section*{References}

\re Becker R.H., White R.L., Edwards A.L. 1991, ApJS 75, 1

\re Browne I.W.A., Patnaik A.R., Wilkinson P.N., Wrobel J.M. 1998, MNRAS 293, 257

\re Chikada et al. 1991, in {\it Frontiers of VLBI}, eds. Hirabayashi et al., Universal Academy Press (Tokyo), p79

\re Douglas J.N., Bash F.N., Bozyan F.A., Torrence G.W., Wolfe C. 1996, AJ 111, 1945

\re Johnston K.J., Fey A.L., Zacharias N., Russell J.L., Ma C., de Vegt C., Reynolds J.E., Jauncey D.L., Archinal B.A., Carter M.S., Corbin T.E., Eubanks T.M., Florkowski D.R., Hall D.M., McCarthy D.D., McCulloch P.M., King E.A., Nicolson G., Shaffer D.B. 1995, AJ 110, 880

\re Griffith M.R., Wright A.E., Burke B.F., Ekers R.D. 1994, ApJS 90, 179

\re Griffith M.R., Wright A.E., Burke B.F., Ekers R.D. 1995, ApJS 97, 347

\re Honma M., Kawaguchi N., Sasao T. 2000, in Proc. of SPIE Vol.4015 {\it Radio Telescope}, eds H. R. Buthcer, in press

\re Kameya O., Sasao T., Kawano N., Hara T., Kuji S., Iwadate K., Tsuruta S., Asari K., Tamura Y., Yasuda S., Morimoto M., Miyazawa K., Kawaguchi N., Miyaji T. 1993, in Proc. of the Second North-Asian Regional Meeting {\it Recent Developement in Millimeter-Wave and Infrared Astronomy}, eds S.H. Cho and H.S. Chung, Terra Scientific Publishing Company (Tokyo) pp16-20

\re Kawaguchi N., Sasao T., Manabe S. 2000, in Proc. of SPIE Vol.4015 {\it Radio Telescope}, eds H. R. Buthcer, in press

\re Lestrade J.F., Preston R.A., Jones D.L., Phillips R.B., Rogers A.E.E., Titus M.A., Rioja M.J., Gabuzda D.C. 1999, A\&A 344, 1014

\re Ma C., Arias E.F., Eubanks T.M., Fey A.L., Gontier A.M., Jacobs C.S., Sovers O.J., Archinal B.A., Charlot P. 1998, AJ 116, 516

\re Morabito D.D., Niell A.E., Preston R.A., Linfield R.P., Wehrle A.E., Faulkner J. 1986, AJ 91, 1038

\re Omodaka T., Morimoto M., Kawaguchi N., Miyaji T., Yasuda S., Suzuyama T., Kitagawa T., Miyazaki T., Furuya L., Jike T., Miyazawa K., Mikoshiba H., Kuji S., Kameya O. 1994, in {\it VLBI TECHNOLOGY - Progress and Future Observational Possibilities}, eds Sasao T., Manabe S., Kameya O., Inoue M., Terra Scientific Publishing Company (Tokyo), pp191-195

\re Palagi F., Cesaroni R.,Comoretto G., Felli M., Natale V. 1993, A\&AS 101, 153

\re Patnaik A.R., Browne I.W.A., Wilkinson P.N., Wrobel J.M., 1992, MNRAS 154, 655

\re Peck A.B., Beasley A.J. 1998, in {\it Radio Emission from Galactic and Extragalactic Compact Sources}, ASP Conference Series Vol.144, IAU Colloquium 464, eds, J.A. Zensus, G.B. Taylor, and J.M. Wrobel, p.155

\re Preston R.A., Morabito D.D., Williams J.G., Faulkner J., Jauncey J., Nicolson G.D. 1985, AJ 90, 1599

\re Shibata K.M., Asaki Y., Asari I., Fukuzaki Y., Hara T., Horiai K., Iwadate K., Kameya O., Kawano N., Kuji S., Manabe S., Sakai S., Sasao T., Sato K., Tamura Y., Tsuruta S. 1994, in {\it VLBI TECHNOLOGY - Progress and Future Observational Possibilities}, eds Sasao T., Manabe S., Kameya O., Inoue M., Terra Scientific Publishing Company (Tokyo), pp185-190

\re Wilkinson P.N., Browne I.W.A., Patnaik A.R., Wrobel J.M.,Sorathia B. 1998, MNRAS 300, 790

\end{document}


\begin{table*}
\begin{center}
Table~1. \hspace{4pt} List of 284 VLBI source candidates.\\
\end{center}
\vspace{6pt}

\begin{tabular*}{\textwidth}{@{\hspace{\tabcolsep}
\extracolsep{\fill}}cllrrrrc}
\hline\hline\\ [-6pt]
Name & RA(J2000) & Dec(J2000) & $l$ $(^\circ$)& $b$ ($^\circ$)& S/N & $S_{\rm 22G}$ & Note \\
\hline\hline\\
TXS2358+605 & 00 01 07.4 & +60 51 25 & 116.84 &  -1.42 &   7.5 &  109 &   \\
TXS0009+655 & 00 12 37.6 & +65 51 11 & 118.97 &   3.28 &   ... &  ... &   \\
TXS0012+610 & 00 14 48.7 & +61 17 45 & 118.55 &  -1.26 &  13.5 &  206 &   \\
TXS0026+637 & 00 29 44.8 & +63 58 43 & 120.56 &   1.21 &   ... &  ... &   \\
TXS0033+586 & 00 36 07.7 & +58 55 47 & 120.95 &  -3.88 &   ... &  ... &   \\
TXS0034+623 & 00 37 04.6 & +62 36 30 & 121.28 &  -0.22 &   ... &  ... &   \\
TXS0034+602 & 00 37 24.2 & +60 29 51 & 121.20 &  -2.33 &   ... &  ... &   \\
TXS0041+660 & 00 44 41.3 & +66 18 42 & 122.25 &   3.45 &   ... &  ... &   \\
TXS0042+672 & 00 45 42.1 & +67 33 21 & 122.38 &   4.69 &   ... &  ... &   \\
TXS0055+675 & 00 58 19.4 & +67 49 01 & 123.58 &   4.95 &   ... &  ... &   \\
TXS0056+579 & 00 59 52.2 & +58 12 24 & 124.05 &  -4.65 &  15.5 &  231 &   \\
TXS0056+665 & 01 00 03.8 & +66 50 58 & 123.78 &   3.99 &   ... &  ... &   \\
TXS0059+581 & 01 02 45.7630 & +58 24 11.139 & 124.42 &  -4.44 &  26.0 &  444 & IV\\
TXS0103+631 & 01 06 29.7 & +63 26 06 & 124.61 &   0.61 &   ... &  ... &   \\
TXS0106+612 & 01 09 46.6 & +61 33 31 & 125.11 &  -1.24 &  16.7 &  219 &   \\
TXS0137+635 & 01 40 43.1 & +63 46 07 & 128.35 &   1.42 &  12.4 &  165 &   \\
TXS0141+579 & 01 45 14.2993 & +58 10 49.783 & 129.99 &  -3.95 &   ... &  ... & V \\
TXS0144+584 & 01 47 46.5438 & +58 40 44.975 & 130.21 &  -3.39 &   ... &  ... & V \\
TXS0203+625 & 02 07 02.7 & +62 46 13 & 131.46 &   1.16 &  17.4 &  226 &   \\
TXS0205+643 & 02 09 36.0 & +64 37 26 & 131.18 &   3.01 &  14.1 &  169 &   \\
TXS0205+614 & 02 09 37.9 & +61 40 55 & 132.06 &   0.20 &   ... &  ... &   \\
TXS0219+628 & 02 23 29.2 & +63 07 17 & 133.12 &   2.08 &   ... &  ... &   \\
TXS0227+626 & 02 31 59.0 & +62 50 36 & 134.12 &   2.18 &   5.6 &   71 &   \\
TXS0233+647 & 02 37 50.4 & +64 57 55 & 133.89 &   4.38 &   ... &  ... &   \\
TXS0237+612 & 02 41 33.8 & +61 26 29 & 135.70 &   1.33 &   ... &  ... &   \\
TXS0241+622 & 02 44 57.6968 & +62 28 06.514 & 135.64 &   2.43 &  47.8 &  799 & I \\
TXS0248+560 & 02 51 54.8 & +56 16 20 & 139.13 &  -2.76 &  35.2 &  495 &   \\
TXS0301+627 & 03 05 53.9 & +62 57 22 & 137.55 &   3.96 &   ... &  ... &   \\
TXS0302+625 & 03 06 42.6611 & +62 43 02.022 & 137.75 &   3.80 &   8.9 &  118 & IV\\
TXS0309+525 & 03 12 50.7 & +52 45 25 & 143.49 &  -4.38 &   ... &  ... &   \\
TXS0314+565 & 03 17 55.1 & +56 43 42 & 142.03 &  -0.61 &   ... &  ... &   \\
TXS0323+551 & 03 27 18.6 & +55 20 28 & 143.89 &  -1.06 &   ... &  ... &   \\
TXS0326+533 & 03 29 52.2 & +53 32 34 & 145.21 &  -2.33 &   ... &  ... &   \\
TXS0331+544 & 03 34 56.0 & +54 37 23 & 145.20 &  -1.02 &   ... &  ... &   \\
TXS0334+506 & 03 37 43.3 & +50 45 53 & 147.81 &  -3.90 &   ... &  ... &   \\
TXS0342+538 & 03 46 32.0 & +54 01 25 & 146.91 &  -0.49 &  13.2 &  177 &   \\
TXS0346+514 & 03 50 24.9 & +51 38 37 & 148.85 &  -1.99 &   ... &  ... &   \\
TXS0346+518 & 03 50 44.4 & +52 01 22 & 148.65 &  -1.66 &   5.1 &   49 &   \\
TXS0354+559 & 03 58 30.0 & +56 06 44 & 146.92 &   2.21 &  30.8 &  436 &   \\
TXS0355+508 & 03 59 29.7477 & +50 57 50.161 & 150.38 &  -1.60 & 330.3 & 3864 & IV\\
TXS0410+486 & 04 14 39.2 & +48 48 14 & 153.64 &  -1.57 &   ... &  ... &   \\
TXS0414+548 & 04 18 19.3405 & +54 57 15.327 & 149.77 &   3.25 &   8.5 &  124 & V \\
TXS0418+532 & 04 22 44.4004 & +53 24 26.240 & 151.32 &   2.61 &  10.1 &  143 & V \\
TXS0424+551 & 04 28 46.3 & +55 18 11 & 150.57 &   4.56 &   ... &  ... &   \\
TXS0428+445 & 04 31 48.1 & +44 42 19 & 158.62 &  -2.36 &   ... &  ... &   \\
TXS0429+415 & 04 32 36.5044 & +41 38 28.433 & 160.97 &  -4.34 &  25.0 &  347 & V \\
TXS0435+487 & 04 38 59.1 & +48 48 46 & 156.42 &   1.32 &   ... &  ... &   \\
TXS0436+426 & 04 40 07.8720 & +42 44 40.251 & 161.08 &  -2.57 &   7.0 &  102 & V \\
TXS0445+449 & 04 48 58.2 & +45 02 01 & 160.39 &   0.13 &   ... &  ... &   \\
TXS0446+422 & 04 49 33.6 & +42 17 46 & 162.56 &  -1.54 &   ... &  ... &   \\
\hline
\end{tabular*}
\end{table*}
\newpage
\begin{table*}
\begin{tabular*}{\textwidth}{@{\hspace{\tabcolsep}
\extracolsep{\fill}}cllrrrrc}
\hline\\
TXS0447+408 & 04 50 43.5 & +40 56 13 & 163.74 &  -2.25 &  13.8 &  152 &   \\
TXS0446+518 & 04 50 53.0 & +51 56 34 & 155.27 &   4.79 &   ... &  ... &   \\
TXS0447+482 & 04 51 44.1 & +48 19 48 & 158.16 &   2.60 &   ... &  ... &   \\
TXS0456+428 & 05 00 27.5 & +42 53 28 & 163.36 &   0.39 &   6.2 &   69 &   \\
TXS0458+476 & 05 02 25.9 & +47 46 05 & 159.71 &   3.65 &   ... &  ... &   \\
TXS0500+443 & 05 03 50.0 & +44 24 38 & 162.53 &   1.80 &   7.1 &   79 &   \\
TXS0500+339 & 05 03 56.7859 & +34 03 28.138 & 170.77 &  -4.47 &   6.1 &   72 & V \\
TXS0501+380 & 05 04 42.3 & +38 05 17 & 167.65 &  -1.91 &   ... &  ... &   \\
TXS0503+466 & 05 07 23.6580 & +46 45 42.349 & 161.03 &   3.72 &  16.0 &  178 & V \\
TXS0509+406 & 05 12 52.5 & +40 41 43 & 166.50 &   0.92 &  28.7 &  431 &   \\
TXS0510+403 & 05 14 03.0 & +40 25 23 & 166.85 &   0.94 &   ... &  ... &   \\
TXS0513+455 & 05 17 28.8994 & +45 37 04.871 & 162.98 &   4.46 &   7.0 &   87 & V \\
TXS0514+330 & 05 18 05.1438 & +33 06 13.369 & 173.28 &  -2.65 &  10.4 &  164 & V \\
TXS0514+374 & 05 18 10.1 & +37 33 38 & 169.65 &  -0.07 &   ... &  ... &   \\
TXS0517+454 & 05 20 47.5 & +45 28 35 & 163.43 &   4.86 &   ... &  ... &   \\
TXS0523+327 & 05 26 43.4 & +32 50 09 & 174.53 &  -1.31 &   ... &  ... &   \\
TXS0526+373 & 05 30 12.7 & +37 23 32 & 171.13 &   1.81 &   6.7 &   91 &   \\
TXS0532+391 & 05 35 55.1 & +39 10 58 & 170.24 &   3.72 &   ... &  ... &   \\
TXS0534+342 & 05 37 45.1 & +34 18 07 & 174.56 &   1.41 &   ... &  ... &   \\
TXS0537+356 & 05 40 52.9 & +35 42 16 & 173.71 &   2.70 &   ... &  ... &   \\
TXS0539+290 & 05 42 11.9 & +29 01 46 & 179.53 &  -0.59 &   ... &  ... &   \\
TXS0544+273 & 05 47 34.1490 & +27 21 56.845 & 181.57 &  -0.45 &  11.8 &  183 &IVM\\
TXS0545+276 & 05 49 06.8 & +27 37 11 & 181.52 &  -0.02 &   ... &  ... &   \\
TXS0547+234 & 05 50 47.3 & +23 26 48 & 185.30 &  -1.84 &  11.4 &  148 &   \\
TXS0549+192 & 05 52 25.8 & +19 13 40 & 189.13 &  -3.66 &   6.5 &   84 &   \\
TXS0550+312 & 05 53 21.6 & +31 17 51 & 178.83 &   2.65 &   ... &  ... &   \\
TXS0550+195 & 05 53 42.1 & +19 36 30 & 188.95 &  -3.21 &   ... &  ... &   \\
TXS0556+238 & 05 59 32.0331 & +23 53 53.926 & 185.92 &   0.12 &  22.5 &  336 & IM\\
TXS0558+193 & 06 01 54.4 & +19 23 16 & 190.11 &  -1.64 &   ... &  ... &   \\
TXS0600+177 & 06 03 09.1302 & +17 42 16.810 & 191.72 &  -2.21 &  14.2 &  165 & IM\\
TXS0600+219 & 06 03 51.5 & +21 59 37 & 188.07 &   0.04 &   6.0 &   75 &   \\
TXS0600+299 & 06 03 55.8520 & +29 57 05.339 & 181.14 &   3.95 &   ... &  ... & V \\
TXS0601+245 & 06 04 55.5 & +24 29 55 & 186.01 &   1.48 &  26.0 &  437 & M \\
TXS0610+171 & 06 13 36.4 & +17 08 25 & 193.43 &  -0.31 &   5.9 &   68 &   \\
TXS0610+260 & 06 13 49.1 & +26 04 37 & 185.59 &   4.00 &  10.4 &  164 & M \\
TXS0611+131 & 06 13 57.6934 & +13 06 45.401 & 197.01 &  -2.16 &  21.6 &  256 & IV\\
TXS0616+136 & 06 18 57.5 & +13 36 32 & 197.15 &  -0.85 &   ... &  ... &   \\
TXS0616+051 & 06 19 00.2 & +05 06 30 & 204.66 &  -4.84 &   ... &  ... &   \\
TXS0617+210 & 06 20 19.5286 & +21 02 29.501 & 190.74 &   2.94 &   ... &  ... & V \\
TXS0618+145 & 06 21 41.1 & +14 32 11 & 196.64 &   0.17 &   ... &  ... &   \\
TXS0622+147 & 06 25 46.0 & +14 40 19 & 196.98 &   1.10 &   5.0 &   59 &   \\
TXS0624+139 & 06 27 35.8 & +13 56 50 & 197.83 &   1.16 &   ... &  ... &   \\
TXS0625+034 & 06 27 38.2 & +03 24 59 & 207.17 &  -3.72 &   ... &  ... &   \\
TXS0626+077 & 06 29 06.7 & +07 43 32 & 203.51 &  -1.40 &   ... &  ... &   \\
TXS0627+176 & 06 30 07.2587 & +17 38 12.930 & 194.84 &   3.40 &   9.7 &  158 & V \\
TXS0628+203 & 06 31 01.0614 & +20 20 59.206 & 192.52 &   4.84 &  14.5 &  221 & V \\
TXS0628+191 & 06 31 31.7 & +19 08 33 & 193.65 &   4.39 &   ... &  ... &   \\
TXS0629+104 & 06 32 15.3 & +10 22 02 & 201.53 &   0.51 &   5.1 &   62 &   \\
TXS0629+160 & 06 32 43.1349 & +15 59 57.606 & 196.58 &   3.20 &  14.0 &  208 &VM \\
TXS0630+082 & 06 33 13.0 & +08 13 20 & 203.54 &  -0.27 &   ... &  ... &   \\
\hline\\
\end{tabular*}
\end{table*}
\newpage
\begin{table*}
\begin{tabular*}{\textwidth}{@{\hspace{\tabcolsep}
\extracolsep{\fill}}cllrrrrc}
\hline\\
TXS0633+045 & 06 36 05.7 & +04 32 41 & 207.14 &  -1.33 &   ... &  ... &   \\
TXS0634+013 & 06 36 56.5 & +01 18 20 & 210.11 &  -2.63 &   ... &  ... &   \\
TXS0637+028 & 06 40 12.9 & +02 50 57 & 209.12 &  -1.19 &   ... &  ... &   \\
TXS0638+098 & 06 41 45.1 & +09 46 48 & 203.13 &   2.32 &   ... &  ... &   \\
TXS0639-032 & 06 41 51.0 & -03 20 48 & 214.82 &  -3.66 &  16.0 &  362 &   \\
TXS0640+090 & 06 43 26.3 & +08 57 38 & 204.05 &   2.31 &  12.5 &  176 &   \\
TXS0641-010 & 06 43 48.1 & -01 08 21 & 213.08 &  -2.22 &   ... &  ... &   \\
TXS0641+127 & 06 43 59.8 & +12 38 18 & 200.83 &   4.10 &   ... &  ... &   \\
TXS0642+055 & 06 45 16.7 & +05 31 33 & 207.31 &   1.15 &   ... &  ... &   \\
TXS0645-023 & 06 48 29.0 & -02 22 07 & 214.70 &  -1.74 &   ... &  ... &   \\
TXS0647-092 & 06 49 24.7 & -09 18 37 & 221.01 &  -4.68 &   ... &  ... &   \\
TXS0647-018 & 06 50 25.7 & -01 52 22 & 214.48 &  -1.08 &   ... &  ... &   \\
TXS0648-083 & 06 50 59.3 & -08 23 52 & 220.37 &  -3.92 &   ... &  ... &   \\
TXS0650-063 & 06 53 00.7 & -06 25 32 & 218.83 &  -2.58 &   7.1 &  308 &   \\
TXS0650-003 & 06 53 17.1 & -00 23 02 & 213.48 &   0.24 &   ... &  ... &   \\
TXS0651+001 & 06 53 50.5 & +00 05 08 & 213.13 &   0.57 &   ... &  ... &   \\
TXS0653-033 & 06 56 11.2 & -03 23 06 & 216.49 &  -0.49 &  17.6 &  576 &   \\
TXS0656-062 & 06 59 03.2 & -06 21 09 & 219.45 &  -1.21 &   ... &  ... &   \\
TXS0700-007 & 07 03 19.1 & -00 51 01 & 215.04 &   2.25 &  28.3 &  390 &   \\
TXS0703-133 & 07 06 00.6 & -13 24 32 & 226.52 &  -2.91 &   --- &  --- &   \\
TXS0704-003 & 07 07 29.9 & -00 26 07 & 215.15 &   3.37 &   ... &  ... &   \\
TXS0705-079 & 07 07 49.6 & -07 59 38 & 221.91 &  -0.03 &   ... &  ... &   \\
TXS0705-142 & 07 08 03.4 & -14 21 15 & 227.59 &  -2.90 &   --- &  --- &   \\
TXS0706-042 & 07 08 38.9 & -04 19 06 & 218.74 &   1.85 &   ... &  ... &   \\
TXS0706-153 & 07 09 12.2 & -15 27 00 & 228.69 &  -3.16 &   --- &  --- &   \\
TXS0707-028 & 07 09 48.4 & -02 56 22 & 217.65 &   2.74 &   ... &  ... &   \\
TXS0707-036 & 07 10 20.2 & -03 43 27 & 218.40 &   2.50 &  11.0 &  133 &   \\
TXS0708+005 & 07 10 53.5 & +00 26 49 & 214.75 &   4.53 &   ... &  ... &   \\
TXS0709+008 & 07 11 47.3 & +00 48 19 & 214.54 &   4.89 &   ... &  ... &   \\
TXS0713-192 & 07 15 39.1 & -19 20 05 & 232.85 &  -3.58 &   --- &  --- &   \\
TXS0713-024 & 07 15 46.9 & -02 31 02 & 217.96 &   4.26 &   ... &  ... &   \\
TXS0714-165 & 07 16 59.6 & -16 35 59 & 230.57 &  -2.03 &   ... &  ... &   \\
TXS0715-181 & 07 18 14.5 & -18 12 59 & 232.14 &  -2.52 &   --- &  --- &   \\
TXS0718-152 & 07 20 44.3 & -15 20 39 & 229.88 &  -0.64 &   ... &  ... &   \\
TXS0718-154 & 07 21 13.6 & -15 30 38 & 230.09 &  -0.62 &   ... &  ... &   \\
TXS0718-119 & 07 21 21.9 & -12 05 33 & 227.09 &   1.02 &   ... &  ... &   \\
TXS0719-187 & 07 22 09.5 & -18 48 30 & 233.10 &  -1.97 &   --- &  --- &   \\
TXS0721-093 & 07 23 27.6 & -09 25 45 & 224.98 &   2.73 &   ... &  ... &   \\
TXS0721-071 & 07 24 17.4 & -07 15 18 & 223.15 &   3.93 &  12.5 &  218 &   \\
TXS0722-068 & 07 24 29.1 & -06 58 08 & 222.92 &   4.11 &   ... &  ... &   \\
TXS0722-095 & 07 24 55.2 & -09 39 57 & 225.36 &   2.93 &   ... &  ... &   \\
TXS0725-255 & 07 27 33.2 & -25 36 31 & 239.70 &  -4.08 &   --- &  --- &   \\
TXS0727-221 & 07 29 26.1 & -22 18 06 & 236.99 &  -2.13 &   --- &  --- &   \\
TXS0727-115 & 07 30 19.1124 & -11 41 12.601 & 227.77 &   3.14 &  68.5 & 1270 &IVM\\
TXS0728-262 & 07 30 42.3 & -26 20 36 & 240.68 &  -3.80 &   --- &  --- &   \\
TXS0728-235 & 07 31 06.6 & -23 41 49 & 238.40 &  -2.46 &   --- &  --- &   \\
TXS0729-223 & 07 31 31.9 & -22 24 40 & 237.32 &  -1.76 &   --- &  --- &   \\
TXS0733-174 & 07 35 45.8125 & -17 35 48.501 & 233.58 &   1.44 &  14.0 &  284 & I \\
TXS0733-224 & 07 35 57.1 & -22 32 59 & 237.93 &  -0.93 &   --- &  --- &   \\
TXS0738-137 & 07 40 20.5 & -13 51 52 & 230.86 &   4.23 &   ... &  ... &   \\
\hline\\
\end{tabular*}
\end{table*}
\newpage
\begin{table*}
\begin{tabular*}{\textwidth}{@{\hspace{\tabcolsep}
\extracolsep{\fill}}cllrrrrc}
\hline\\
TXS0739-266 & 07 41 55.4 & -26 47 30 & 242.30 &  -1.83 &   --- &  --- &   \\
TXS0740-171 & 07 42 25.4 & -17 15 11 & 234.06 &   3.00 &   ... &  ... &   \\
TXS0741-249 & 07 43 21.9 & -25 03 09 & 240.95 &  -0.69 &   --- &  --- &   \\
TXS0741-183 & 07 43 32.2 & -18 25 57 & 235.22 &   2.64 &   ... &  ... &   \\
TXS0744-158 & 07 46 18.5 & -15 55 36 & 233.37 &   4.46 &   5.7 &  112 &   \\
TXS0745-191 & 07 47 31.4 & -19 17 39 & 236.44 &   3.03 &   --- &  --- &   \\
TXS0745-165 & 07 48 03.1 & -16 39 50 & 234.22 &   4.46 &  13.7 &  276 &   \\
TXS0749-191 & 07 51 23.1 & -19 17 17 & 236.90 &   3.82 &   --- &  --- &   \\
TXS0750-219 & 07 52 35.9 & -22 02 40 & 239.42 &   2.66 &   --- &  --- &   \\
TXS1729-244 & 17 32 05.7 & -24 26 52 &   2.20 &   4.95 &   ... &  ... &   \\
TXS1744-193 & 17 47 14.4 & -19 20 59 &   8.40 &   4.65 &   ... &  ... &   \\
TXS1748-253 & 17 51 51.2630 & -25 24 00.060 &   3.74 &   0.63 &   ... &  ... & I \\
TXS1749-224 & 17 52 33.2 & -22 30 12 &   6.32 &   1.97 &   ... &  ... &   \\
TXS1750-187 & 17 53 09.0 & -18 43 35 &   9.65 &   3.77 &   ... &  ... &   \\
TXS1752-225 & 17 55 26.0 & -22 32 09 &   6.63 &   1.38 &   ... &  ... &   \\
TXS1754-226 & 17 57 27.7 & -22 38 59 &   6.77 &   0.92 &   ... &  ... &   \\
TXS1755-162 & 17 58 41.5 & -16 16 52 &  12.44 &   3.84 &   ... &  ... &   \\
TXS1755-179 & 17 58 43.3 & -17 57 51 &  10.98 &   3.00 &   ... &  ... &   \\
TXS1759-224 & 18 02 13.1 & -22 27 51 &   7.47 &   0.06 &   ... &  ... &   \\
TXS1800-218 & 18 03 05.0 & -21 49 31 &   8.13 &   0.20 &   ... &  ... &   \\
TXS1800-278 & 18 03 16.7 & -27 48 09 &   2.94 &  -2.78 &   ... &  ... &   \\
TXS1800-175 & 18 03 38.1 & -17 32 25 &  11.92 &   2.19 &   ... &  ... &   \\
TXS1802-252 & 18 05 23.5 & -25 12 38 &   5.44 &  -1.92 &   ... &  ... &   \\
TXS1806-200 & 18 09 00.8 & -20 04 51 &  10.33 &  -0.16 &   ... &  ... &   \\
TXS1808-194 & 18 11 35.3 & -19 26 04 &  11.18 &  -0.37 &   ... &  ... &   \\
TXS1811-179 & 18 14 15.1 & -17 55 51 &  12.81 &  -0.21 &   ... &  ... &   \\
TXS1811-167 & 18 14 35.4 & -16 45 32 &  13.87 &   0.28 &   ... &  ... &   \\
TXS1815-111 & 18 18 16.4 & -11 08 09 &  19.25 &   2.17 &   ... &  ... &   \\
TXS1816-150 & 18 18 58.8 & -15 00 28 &  15.92 &   0.19 &   ... &  ... &   \\
TXS1817-044 & 18 20 13.7 & -04 25 42 &  25.41 &   4.89 &   ... &  ... &   \\
TXS1817-098 & 18 20 38.1 & -09 47 17 &  20.71 &   2.29 &   ... &  ... &   \\
TXS1819-245 & 18 22 21.1 & -24 30 03 &   7.90 &  -4.97 &   ... &  ... &   \\
TXS1819-096 & 18 22 28.8 & -09 38 57 &  21.05 &   1.96 &   ... &  ... &   \\
TXS1822-173 & 18 25 36.6 & -17 18 49 &  14.63 &  -2.30 &   7.2 &  130 &   \\
TXS1822-076 & 18 25 37.8 & -07 37 28 &  23.20 &   2.21 &   6.6 &  131 &   \\
TXS1825-041 & 18 27 45.0411 & -04 05 44.568 &  26.58 &   3.38 &   ... &  ... & V \\
TXS1825-214 & 18 28 19.5 & -21 23 37 &  11.30 &  -4.76 &   ... &  ... &   \\
TXS1825-055 & 18 28 39.8 & -05 30 50 &  25.42 &   2.53 &   ... &  ... &   \\
TXS1829-106 & 18 32 20.8 & -10 35 12 &  21.35 &  -0.63 &  29.0 &  564 &   \\
TXS1830-106 & 18 33 33.6 & -10 34 06 &  21.50 &  -0.89 &   ... &  ... &   \\
TXS1831-073 & 18 34 10.3 & -07 17 52 &  24.47 &   0.49 &   ... &  ... &   \\
TXS1835-055 & 18 38 16.4 & -05 29 46 &  26.54 &   0.41 &   ... &  ... &   \\
TXS1836+040 & 18 38 48.8289 & +04 04 24.637 &  35.13 &   4.66 &  11.4 &  191 & V \\
TXS1836-125 & 18 39 30.4 & -12 30 55 &  20.44 &  -3.07 &   ... &  ... &   \\
TXS1840-079 & 18 42 48.6 & -07 56 06 &  24.89 &  -1.71 &   ... &  ... &   \\
TXS1840-036 & 18 43 16.9 & -03 35 38 &  28.80 &   0.17 &   ... &  ... &   \\
TXS1841-043 & 18 44 22.4 & -04 17 43 &  28.30 &  -0.39 &   ... &  ... &   \\
TXS1843-001 & 18 46 03.9 & -00 03 38 &  32.27 &   1.17 &   5.6 &   76 &   \\
TXS1843-120 & 18 46 06.3 & -11 58 09 &  21.66 &  -4.26 &   ... &  ... &   \\
TXS1843-030 & 18 46 27.2 & -02 59 11 &  29.71 &  -0.25 &   ... &  ... &   \\
\hline\\
\end{tabular*}
\end{table*}
\newpage
\begin{table*}
\begin{tabular*}{\textwidth}{@{\hspace{\tabcolsep}
\extracolsep{\fill}}cllrrrrc}
\hline\\
TXS1844-025 & 18 47 00.5 & -02 27 52 &  30.23 &  -0.14 &   ... &  ... &   \\
TXS1844-001 & 18 47 04.5 & -00 04 48 &  32.36 &   0.93 &   ... &  ... &   \\
TXS1845+018 & 18 47 44.0 & +01 54 33 &  34.21 &   1.69 &   ... &  ... &   \\
TXS1847-092 & 18 50 29.7 & -09 08 54 &  24.67 &  -3.95 &   ... &  ... &   \\
TXS1857+129 & 18 59 23.3 & +12 58 55 &  45.40 &   4.15 &   ... &  ... &   \\
TXS1901+155 & 19 04 14.3609 & +15 36 38.448 &  48.28 &   4.29 &   7.8 &  109 & V \\
TXS1902+159 & 19 04 46.2 & +16 03 15 &  48.74 &   4.38 &   ... &  ... &   \\
TXS1903+015 & 19 06 13.0 & +01 39 13 &  36.09 &  -2.53 &   ... &  ... &   \\
TXS1904+013 & 19 07 11.9969 & +01 27 08.974 &  36.02 &  -2.84 &   7.8 &  117 & V \\
TXS1906+165 & 19 09 01.0 & +16 39 45 &  49.75 &   3.75 &   ... &  ... &   \\
TXS1909+048 & 19 11 49.5 & +04 58 57 &  39.69 &  -2.24 &   ... &  ... &   \\
TXS1910+052 & 19 12 54.2 & +05 18 01 &  40.10 &  -2.34 &  13.4 &  170 &   \\
TXS1911+165 & 19 14 14.4 & +16 36 39 &  50.28 &   2.62 &   7.7 &   96 &   \\
TXS1915+062 & 19 17 48.6 & +06 21 22 &  41.60 &  -2.93 &   ... &  ... &   \\
TXS1916+062 & 19 19 17.3 & +06 19 42 &  41.75 &  -3.27 &   6.5 &   81 &   \\
TXS1919+086 & 19 22 18.6 & +08 41 57 &  44.20 &  -2.82 &   ... &  ... &   \\
TXS1920+154 & 19 22 34.6995 & +15 30 10.024 &  50.23 &   0.33 &  17.2 &  219 & V \\
TXS1921+174 & 19 23 33.7 & +17 29 37 &  52.10 &   1.06 &   ... &  ... &   \\
TXS1922+204 & 19 24 36.5 & +20 35 22 &  54.95 &   2.30 &   ... &  ... &   \\
TXS1922+155 & 19 24 39.4562 & +15 40 43.931 &  50.63 &  -0.03 &  21.9 &  307 & V \\
TXS1923+151 & 19 25 24.5 & +15 14 17 &  50.32 &  -0.40 &   ... &  ... &   \\
TXS1923+210 & 19 25 59.6054 & +21 06 26.162 &  55.56 &   2.26 &  40.3 &  624 & IV\\
TXS1926+273 & 19 28 31.1 & +27 26 58 &  61.43 &   4.76 &   ... &  ... &   \\
TXS1926+087 & 19 28 40.8554 & +08 48 48.412 &  45.04 &  -4.15 &   7.4 &   89 & I \\
TXS1927+256 & 19 29 44.9170 & +25 43 16.223 &  60.03 &   3.70 &   5.6 &   68 & V \\
TXS1928+154 & 19 30 52.8 & +15 32 35 &  51.22 &  -1.41 &  25.0 &  323 &   \\
TXS1929+226 & 19 31 24.9167 & +22 43 31.258 &  57.58 &   1.93 &   9.1 &  120 & I \\
TXS1930+258 & 19 32 49.9 & +25 59 05 &  60.60 &   3.21 &   ... &  ... &   \\
TXS1930+131 & 19 33 02.5 & +13 13 40 &  49.44 &  -2.98 &   ... &  ... &   \\
TXS1931+149 & 19 33 21.8 & +15 04 46 &  51.10 &  -2.16 &   5.4 &   77 &   \\
TXS1932+174 & 19 34 50.2 & +17 32 15 &  53.43 &  -1.28 &   ... &  ... &   \\
TXS1937+214 & 19 39 35.6 & +21 36 14 &  57.52 &  -0.27 &   ... &  ... &   \\
TXS1938+273 & 19 40 39.4 & +27 30 37 &  62.78 &   2.43 &   ... &  ... &   \\
TXS1943+228 & 19 46 06.2519 & +23 00 04.411 &  59.48 &  -0.88 &   5.7 &   71 & I \\
TXS1944+250 & 19 46 47.2 & +25 12 42 &  61.47 &   0.09 &   ... &  ... &   \\
TXS1945+267 & 19 47 26.3 & +26 53 47 &  63.00 &   0.82 &   ... &  ... &   \\
TXS1945+241 & 19 48 01.4 & +24 14 49 &  60.78 &  -0.63 &   ... &  ... &   \\
TXS1950+253 & 19 52 48.1 & +25 26 51 &  62.37 &  -0.96 &  10.7 &  174 &   \\
TXS1954+282 & 19 56 46.5 & +28 20 39 &  65.30 &  -0.22 &  10.1 &  152 &   \\
TXS1958+257 & 20 01 05.4 & +25 51 46 &  63.69 &  -2.34 &   ... &  ... &   \\
TXS1959+241 & 20 01 54.0 & +24 16 20 &  62.44 &  -3.33 &   ... &  ... &   \\
TXS2002+335 & 20 04 23.3 & +33 39 06 &  70.67 &   1.19 &   ... &  ... &   \\
TXS2005+403 & 20 07 44.9449 & +40 29 48.604 &  76.82 &   4.30 &  63.5 &  946 & I \\
TXS2005+372 & 20 07 45.2 & +37 22 01 &  74.18 &   2.61 &   7.8 &  105 &   \\
TXS2008+332 & 20 10 49.7 & +33 22 13 &  71.16 &  -0.09 &  11.4 &  210 &   \\
TXS2013+340 & 20 15 28.8318 & +34 10 39.406 &  72.37 &  -0.45 &  24.0 &  331 & V \\
TXS2014+358 & 20 16 45.7 & +36 00 33 &  74.04 &   0.36 &   6.8 &   94 &   \\
TXS2015+336 & 20 17 13.1 & +33 45 46 &  72.23 &  -0.98 &   ... &  ... &   \\
TXS2018+295 & 20 20 06.5 & +29 42 14 &  69.21 &  -3.77 &   9.2 &  141 &   \\
TXS2018+282 & 20 20 45.8719 & +28 26 59.205 &  68.25 &  -4.59 &   ... &  ... & V \\
\hline\\
\end{tabular*}
\end{table*}
\newpage
\begin{table*}
\begin{tabular*}{\textwidth}{@{\hspace{\tabcolsep}
\extracolsep{\fill}}cllrrrrc}
\hline\\
TXS2021+317 & 20 23 19.0177 & +31 53 02.307 &  71.40 &  -3.10 &  42.3 &  596 & IV\\
TXS2023+335 & 20 25 10.8416 & +33 43 00.218 &  73.13 &  -2.37 & 103.1 & 1496 & IV\\
TXS2023+372 & 20 25 23.5 & +37 23 14 &  76.15 &  -0.28 &   ... &  ... &   \\
TXS2024+296 & 20 26 59.3 & +29 50 53 &  70.18 &  -4.91 &   ... &  ... &   \\
TXS2025+360 & 20 27 22.2 & +36 12 14 &  75.42 &  -1.29 &   ... &  ... &   \\
TXS2027+464 & 20 29 18.9366 & +46 36 02.269 &  84.08 &   4.48 &   ... &  ... & V \\
TXS2048+361 & 20 50 02.2848 & +36 19 52.500 &  78.29 &  -4.85 &  15.8 &  267 & V \\
TXS2056+445 & 20 58 28.6 & +44 46 17 &  85.80 &  -0.65 &   ... &  ... &   \\
TXS2117+494 & 21 18 52.4 & +49 36 58 &  91.71 &   0.08 &   ... &  ... &   \\
TXS2118+443 & 21 20 31.7734 & +44 34 34.281 &  88.32 &  -3.66 &   9.3 &  139 & V \\
TXS2121+547 & 21 23 05.3161 & +55 00 27.329 &  95.98 &   3.44 &   7.8 &  104 & V \\
TXS2122+519 & 21 24 34.8 & +52 07 15 &  94.11 &   1.22 &   ... &  ... &   \\
TXS2125+552 & 21 27 32.5 & +55 28 33 &  96.76 &   3.33 &   ... &  ... &   \\
TXS2126+489 & 21 27 57.6 & +49 10 31 &  92.45 &  -1.27 &   ... &  ... &   \\
TXS2130+556 & 21 32 09.5 & +55 52 41 &  97.51 &   3.17 &   ... &  ... &   \\
TXS2135+508 & 21 37 00.9 & +51 01 34 &  94.77 &  -0.90 &  13.6 &  199 &   \\
TXS2144+568 & 21 46 25.9398 & +57 03 24.687 &  99.77 &   2.75 &   ... &  ... & V \\
TXS2154+482 & 21 56 37.8 & +48 30 28 &  95.58 &  -4.86 &   ... &  ... &   \\
TXS2156+596 & 21 58 29.5 & +59 51 19 & 102.76 &   3.94 &   ... &  ... &   \\
TXS2159+505 & 22 01 43.5394 & +50 48 56.393 &  97.65 &  -3.53 &  16.3 &  260 & V \\
TXS2222+600 & 22 24 36.9 & +60 16 16 & 105.67 &   2.42 &  10.9 &  142 &   \\
TXS2229+539 & 22 31 18.8 & +54 09 36 & 103.26 &  -3.27 &   ... &  ... &   \\
TXS2238+610 & 22 40 20.0 & +61 17 09 & 107.84 &   2.32 &   ... &  ... &   \\
TXS2244+635 & 22 46 13.8 & +63 49 55 & 109.65 &   4.24 &   ... &  ... &   \\
TXS2247+596 & 22 49 28.7 & +59 54 57 & 108.20 &   0.58 &   ... &  ... &   \\
TXS2250+644 & 22 52 05.5 & +64 40 12 & 110.60 &   4.70 &   ... &  ... &   \\
TXS2252+576 & 22 54 53.8 & +57 56 39 & 107.95 &  -1.50 &   ... &  ... &   \\
TXS2255+570 & 22 57 19.8 & +57 20 33 & 107.99 &  -2.18 &   6.5 &   80 &   \\
TXS2259+568 & 23 01 26.6287 & +57 06 25.508 & 108.39 &  -2.63 &  14.6 &  179 & V \\
TXS2300+577 & 23 03 04.8 & +58 03 23 & 108.98 &  -1.85 &   ... &  ... &   \\
TXS2303+599 & 23 05 10.3 & +60 14 39 & 110.11 &   0.05 &   ... &  ... &   \\
TXS2307+627 & 23 09 21.1 & +63 01 03 & 111.65 &   2.40 &   ... &  ... &   \\
TXS2336+598 & 23 39 21.2 & +60 10 13 & 114.08 &  -1.46 &   9.6 &  153 &   \\
TXS2348+643 & 23 50 54.8 & +64 40 20 & 116.51 &   2.56 &   ... &  ... &   \\
\hline
\end{tabular*}
\vspace{6pt}\par\noindent

Column 1: Texas source name; Column 2 \& 3: Source position in J2000; Column 4 \& 5: Source position in Galactic coordinates; Column 6: Signal-to-noise ratio with Nobeyama-Kashima baseline; Column 7: correlated flux density at 22 GHz for Nobeyama-Kashima baseline. {`}---{'} in column 6 \& 7 indicates that the source was not observed.; Column 8: {`}I{'} denotes the source listed in ICRF catalog (Ma et al.1998), {`}V{'} denotes the source listed in VLBA calibrator source (Peck \& Beasley 1998), and {`}M{'} denotes the source listed in 2 GHz VLBI survey (Morabito et al. 1986).
\end{table*}